\documentclass{emulateapj}

\usepackage[latin1]{inputenc}
\usepackage[english]{babel}
\usepackage{amsmath}
\usepackage{amssymb}		% Advanced mathematics
\usepackage{latexsym} 
\usepackage{epsfig}
\usepackage{subfigure}
\usepackage{natbib}
\usepackage{setspace}
\usepackage{graphicx}
\usepackage{longtable}
\newcommand{\msun}{{M_{\odot}}}
\newcommand{\mstar}{{M_{\ast}}}
\newcommand{\ser}{S\'ersic }

\shorttitle{Evolution of the mass-size relation From $z=1-7$}
\shortauthors{Mosleh et al.}

\begin{document}

\title{The Evolution of Mass-size Relation for Lyman Break Galaxies From $z=1$ to $z=7$ }

\author{Moein Mosleh \altaffilmark{1}, Rik J. Williams \altaffilmark{2}, Marijn Franx \altaffilmark{1}, Valentino Gonzalez \altaffilmark{3}, Rychard J. Bouwens \altaffilmark{1}, Pascal Oesch \altaffilmark{3}, Ivo Labbe \altaffilmark{1}, Garth D. Illingworth \altaffilmark{3}, Michele Trenti \altaffilmark{4}}

\altaffiltext{1}{Leiden Observatory, Universiteit Leiden, 2300 RA Leiden, The Netherlands}\email{mosleh@strw.leidenuniv.nl}
\altaffiltext{2}{Carnegie Observatories, Pasadena, CA 91101, USA}
\altaffiltext{3}{UCO/Lick Observatory, University of California, Santa Cruz, CA 95064, USA}
\altaffiltext{4}{Institute of Astronomy, University of Cambridge, Madingley Road, Cambridge CB3 0HA, UK}

\begin{abstract}
For the first time, we study the evolution of the stellar mass-size relation for star-forming galaxies from $z\sim4$ to $z\sim7$ from Hubble-WFC3/IR camera observations of the HUDF and Early Release Science (ERS) field.  The sizes are measured by determining the best fit model to galaxy images in the rest-frame 2100 \AA \ with the stellar masses estimated from SED fitting to rest-frame optical (from Spitzer/IRAC) and UV fluxes. We show that the stellar mass-size relation of Lyman-break galaxies (LBGs) persists, at least to $z\sim5$, and the median size of LBGs at a given stellar mass increases towards lower redshifts. For galaxies with stellar masses of $9.5<\log(\mstar/\msun)<10.4 $ sizes evolve as $(1+z)^{-1.20\pm0.11}$. This evolution is very similar for galaxies with lower stellar masses of $8.6<\log(\mstar/\msun)<9.5$ which is $r_{e} \propto (1+z)^{-1.18\pm0.10}$, in agreement with simple theoretical galaxy formation models at high $z$. Our results are consistent with previous measurements of the LBGs mass-size relation at lower redshifts ($z\sim1-3$).\\

\end{abstract}


\section{Introduction}

The size of a galaxy is a fundamental and important parameter to measure. Over the past decade,  observations have revealed that sizes of galaxies at a given stellar mass were smaller at higher redshifts and change significantly with redshift. It has been shown that the sizes of galaxies correlate with their stellar masses and that this correlation exists at least up to $z\sim3$ \citep[e.g.,][]{franx2008, williams2009, mosleh2011, law2011}. 

There are many proposed scenarios to explain the physical processes of galaxy assembly that plausibly reproduce the observable stellar mass and size of galaxies at different redshifts (e.g., galaxy minor or major mergers \citep[]{khochfar2006,khochfar2009, bell2006, naab2009} or gas accretion in outer regions and star formation \citep[]{dekel2009,Elmegreen2008}. Accurate measurements of both \textit{stellar masses} and \textit{sizes} of galaxies over a wide redshift range are fundamentally important to constrain these galaxy formation models. 

To extend the mass-size relation of galaxies to the highest redshifts, we exploit the Lyman-break galaxies (LBGs) which are star forming galaxies with strong rest-frame UV emission and could be selected by photometric dropout techniques \citep[e.g.][]{steidel2003, adelberger2004} . These galaxies can be identified out to very high redshifts (e.g., $z\sim8$)\citep[]{oesch2012, yan2012} and thus provide insight into the early evolution of the mass-size relation.

Morphological studies of LBGs ($z\sim2-6$) in rest-frame UV have shown that these galaxies are mostly compact sources however, multiple core LBGs have also been found \citep[e.g.][]{ravindranath2006, law2007,conselice2009}. Analysing their size and structure could help to interpret the dominant mechanism for galaxy growth. The new Wide Field Camera 3 (WFC3) on board the Hubble Space Telescope (HST) can provide sizes of high-z galaxies in longer rest-frame wavelengths than Advanced Camera for Survey (ACS).

Size studies of galaxies at redshifts $z>4$ using profile fitting techniques are rare \citep{oesch2010b}. Here for the first time, we investigate the mass-size relation of dropout galaxies up to the very early stages of galaxy formation, using the advantages of wide-area HST surveys and high spatial resolution of WFC3. We measure the sizes of LBGs at approximately the same rest-frame wavelength at different redshifts, minimizing the effects of morphological K-correction. We utilize observations of both Hubble Ultra Deep Field (HUDF) and the deep wide-area Early Release Science (ERS) field to study the mass-size relation of the largest sample of LBGs at $z\sim4-7$ so far.  The cosmological parameters adopted throughout this paper are $\Omega_{m}$ = 0.3, $\Omega_{\Lambda}$ = 0.7 and $H_{0} = 70$ $km$ $s^{-1}$ $Mpc^{-1}$.\\

\begin{figure}
\includegraphics[width=3.4in]{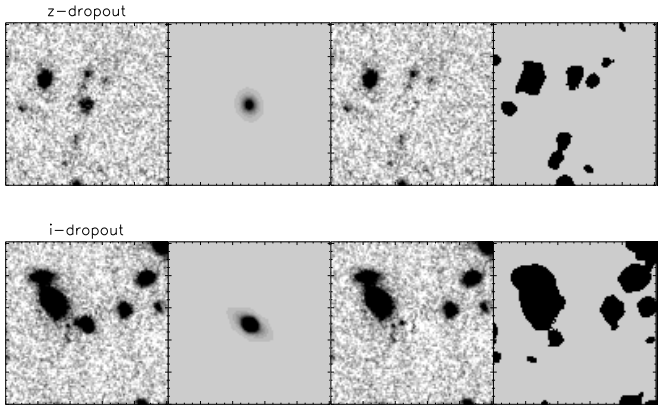}
\caption{From left to right, H-band postage stamps ($6''$ x $6''$), best-fit models from GALFIT, residual images and mask maps are shown for a z-dropout candidate (top row) and an i-dropout candidate (bottom row). The quality of the fits can be seen from the residual images.}
\label{fig1}
\end{figure}

%%%%%%%%%%%%%%%%%%%%%%%%%%%%%%%%%%%%%%%%%%%%%%%%%%
\section{Description of Data}
The sample of dropout sources at $z\sim4-7$ used here is based on recent HST-WFC3/IR observations over the HUDF and ERS field \citep[see][]{bouwens2010, oesch2010a,gonzalez2010}. Our sample of dropout galaxies  is taken from \cite{bouwens2010} and \cite{gonzalez2011}. Their candidates are selected by utilizing color criteria (Lyman-Break technique) for selecting $B$, $V$, $i$ and $z$ dropouts similar to those used in \cite{oesch2010a, bouwens2007, bouwens2010}. Their sample consists of 679 objects at $ z\sim4-6$ in ERS field and 345 dropouts at $z\sim4-7$ in HUDF. 

For all candidate galaxies in this paper we use the stellar masses measured by \cite{gonzalez2011}. These authors used FAST \citep{Kriek2009} to fit template galaxy SEDs to photometry in ACS, WFC3/IR and IRAC $[3.6]$ and $[4.5]$ filters. They used \cite{BC2003} stellar population evolution models with constant star formation histories and assumed a \cite{salpeter1955} initial mass function (IMF; $0.1-100 \msun$) and $0.2 Z_{\sun}$ solar metallicity.  Reliable deblended IRAC fluxes and photometry in different bands (with different PSFs) were obtained by using a source-fitting algorithm described in \cite{labbe2006} \citep[see also][]{gonzalez2010}.
            
The photometric deblending for the IRAC imaging does not properly work for objects in crowded regions due to the large PSF of IRAC bands; due to this, the number of sources with reliable mass estimates is reduced to 60\% \citep[see also][]{gonzalez2011}. The stellar masses are corrected to a \cite{kroupa2001} IMF by a reduction of 0.2 dex \citep[see ][]{Marchesini2009}. These are consistent with the SDSS masses.

The near-IR images used in this study are taken from the full two-year WFC3/IR HUDF ($Y_{105}$,$J_{125}$, $H_{160}$;\citep{bouwens2010, oesch2010a} ) and ERS ($Y_{098}$,$J_{125}$, $H_{160}$; \citep{bouwens2010}) data.  The data set covers $\sim 4.7$ arcmin$^{2}$ in the HUDF and $\sim 40$ arcmin$^{2}$ in the ERS (in GOODS-South) field with a pixel scale of $0.06''$ and a point spread function (PSF) FWHM of $\sim0.17''$ in the H-band. We also use very deep images of the HUDF field obtained with the  ACS \citep{beckwith2006} and deep GOODS ACS/WFC data over the GOODS field \citep{giavalisco2004} with a pixel scale of $0.03''$ and a PSF FWHM of $\sim0.10''$. The reduction of these data is described in \cite{bouwens2010} and references there in.

As explained in the next section, we use sizes of objects in the band closest to the rest-frame 2100\AA. Therefore, in addition to reliable stellar mass cut, we restrict our size measurements to objects with sufficient $S/N$ in their size measurement band. Hence, we use $H_{160}<28.5$ in HUDF and $H_{160}<26.2$ AB mag in ERS field (corresponding to $10\sigma$ in $0.5''$ apertures) for i-dropouts and z-dropouts, $J_{125}<28.5$ in HUDF and $J_{125}<26.5$ AB mag in ERS for V-dropouts and $Y_{105}<28.3$ in HUDF and $Y_{098}<26.4$ AB mag in ERS for B-dropouts.  Our simulations (see Section 3) show that systematic uncertainties on size estimates are small up to these magnitudes. Moreover, objects with poor S/N ($<5\sigma$) are not included in our analysis. Thus, the final sample consists of 156 B-dropouts ($z\sim4$), 45 V-dropouts ($z\sim5$), 13 i-dropouts ($z\sim6$) and 4 z-dropouts ($z\sim7$).\\

\begin{figure*}
\includegraphics[width=\textwidth]{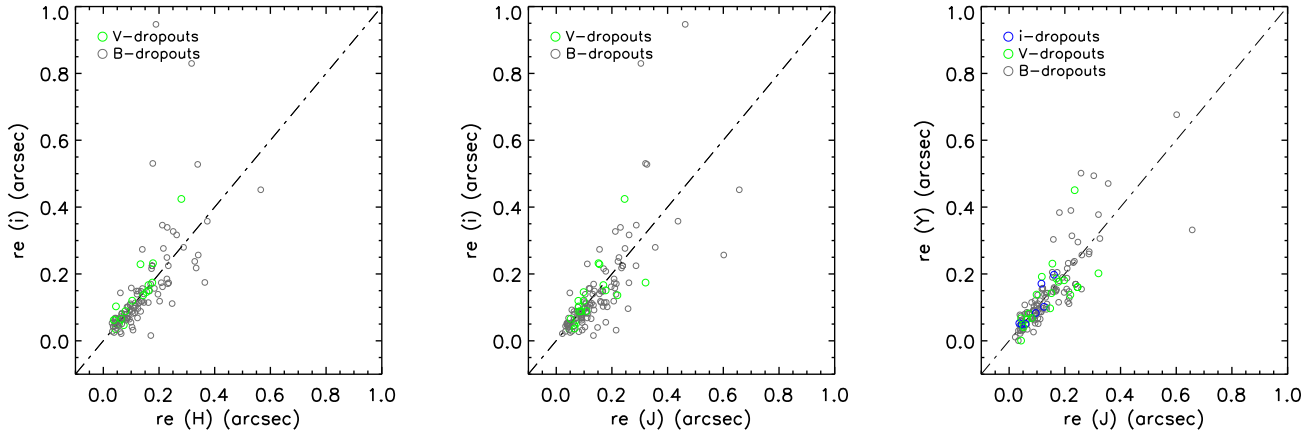}
\caption{Comparison of sizes of dropout candidates measured in different filters with $S/N>10$ in both the HUDF and ERS. Each color represents a different sample of dropout candidates. The correlation between estimated sizes in different bands is an indication of the reliability of our size measurements. The small offsets between sizes in different bands maybe due to color gradients in the galaxies. The objects with large offsets from the one-to-one relation have generally a clumpy structure or multiple cores.}
\label{fig2}
\end{figure*}

%%%%%%%%%%%%%%%%%%%%%%%%%%%%%%%%%%%%%%%%%%%%%%%%%%
\section{Sizes}

Our size measurements are performed in a similar way as in \cite{mosleh2011}. PSF-convolved \ser models are fit to all galaxies using the GALFIT version 3 package \citep{peng2010}. We measure sizes in filters described in the previous section. 

GALFIT requires an accurate point spread function (PSF) of the image to convolve the \ser profile models in order to find the best ($\chi ^{2}$ minimized) fit. For the HUDF WFC3/IR images we use a PSF created using TinyTim \citep{krist1995}. Although, we prefer using empirically-determined PSFs (from stars), in the HUDF there are not enough stars to create a sufficiently low-noise PSF for deconvolution. We note that our test shows that sizes measured using non-saturated star as a PSF give results which are consistent with those based on TinyTim created PSFs. For the ACS images in the HUDF we use non-saturated stars in the field to make a median-stacked PSF. In the ERS field, we use a median stacked-PSF using non-saturated stars in the field for all optical and near-IR bands (ACS and WFC3/IR). 

GALFIT measures the half-light radius along the semimajor axis, $a$, and the axis ratio, $b/a$, of each galaxy. We determine the circularized effective radius of the galaxies ($r_{e}=a\sqrt{b/a}$) to remove the effects of ellipticity. We fix the \ser index to 2.5 where GALFIT fails to return reliable measurements; i.e., for objects with large uncertainties in their output parameters. Size determination for all objects while fixing the \ser index to 2.5 reveals that this will not introduce systematics in our size study. Fixing \ser index to 1.5 or 3 gives similar size estimates as for $n=2.5$. In Figure \ref{fig1} we show from left to right, postage stamps in the H-band, best fit models from GALFIT, residual images and mask maps for a z-dropout (top row) and an i-dropout (bottom row) in the HUDF. The low level residuals shows that our best  fit models closely match the observed galaxy images. 

We first test the accuracy of our measurements by comparing the effective radii, $r_{e}$, measured at different wavelengths. A comparison of the sizes of our dropout candidates is shown in Figure \ref{fig2}. Each dropout sample is represented by a different color: B-dropouts are black, V-dropouts are green and i-dropouts are blue. We note that ACS has a much smaller PSF and finer pixel scale, and it is known to produce highly reliable size measurements from previous studies.  The correlation between estimated sizes in different bands, is an indicator of the size measurements reliability. Since color gradients and clumpy star formation may introduce additional $r_{e}$ scatter between bands, these comparisons essentially show an ``upper limit'' to the intrinsic uncertainties.

In addition, we have performed realistic simulations on both the HUDF and ERS fields analogous to \cite{mosleh2011}, by adding model galaxies to the images and measuring their parameters in the same way as for real objects. From the simulations, the relative difference between the recovered and input sizes versus magnitude shows that the systematic uncertainties on the recovered sizes are very small ($<10\%$) down to S/N ratio of 10  (corresponding to $H_{160}<28.5$ in the HUDF and $H_{160}<26.2$ in the ERS). 
       
Finally, in order to minimize the effects of morphological \textit{K}-correction, we use sizes of each object in the WFC3/IR or ACS band that is closest to the rest-frame 2100\AA. Therefore in this study the $Y_{105}$ band images are used for B dropouts, $J_{125}$ images are used for V dropouts and $H_{160}$ band images for i and z dropouts. \\

%%%%%%%%%%%%%%%%%%%%%%%%%%%%%%%%%%%%%%%%%%%%%%%%%%%
\section{Results}

\subsection{Size-mass Evolution}

\begin{figure*}
\includegraphics[width=\textwidth]{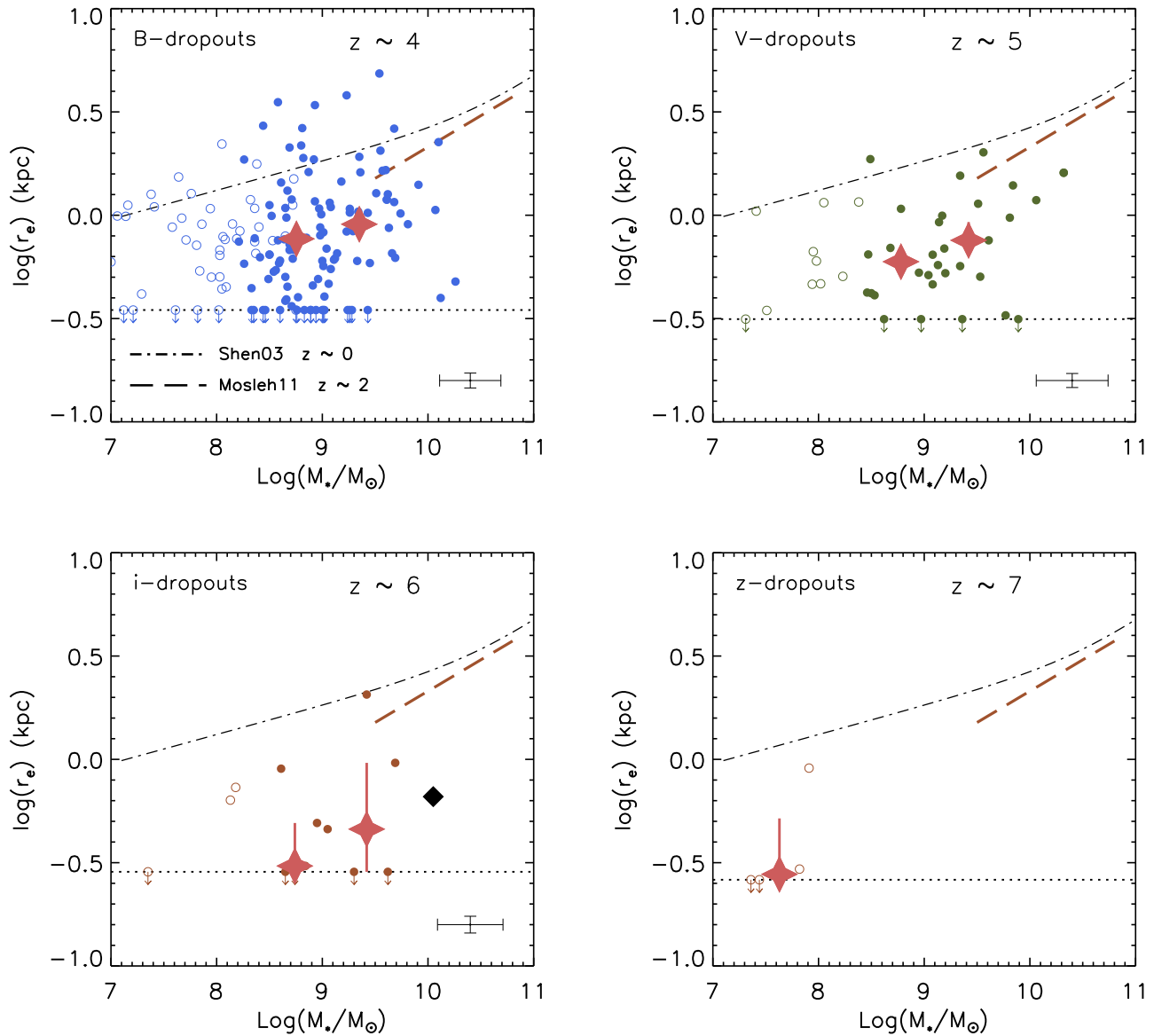}
\caption{The mass-size relation for dropout galaxies in different redshift bins. The filled circles indicate sizes for galaxies with reliable stellar masses ($>2\sigma$ detection in IRAC[3.6]). The pink large stars are the median of sizes from GALFIT in different stellar mass bins with the errors calculated by bootstraping.The  black diamond in the lower left panel is an i-dropout galaxy at $z\sim6$ with a stellar mass of a few times $10^{10}\msun$ from \cite{eyles2005}. The dashed and dot-dashed lines show the mass-size relation for UV-bright galaxies at $z\sim2$ from \cite{mosleh2011} and for late-type galaxies at $z\sim0$ from \cite{shen2003}, respectively. The dotted lines correspond to our size measurement limits. Typical errors of the points are shown in the lower right of each panel. This plot shows that mass-size relation holds out to $z\sim5$, and possibly beyond for galaxies with  $\log(\mstar/\msun)>8.6$. Sizes decrease toward higher redshift. }
\label{fig3}
\end{figure*}

The size and stellar mass estimates of our dropout candidates allow us to investigate the mass-size relation and its evolution. Figure \ref{fig3} shows the mass-size relation for four redshift bin from $z=4$ to $z=7$. There is a clear relation out to $z\sim5$ and a hint at $z\sim6$ for galaxies with  $\log(\mstar/\msun)>8.6$. In each panel, color points represent the distributions of candidates in the mass-size plane. Objects with S/N $<2\sigma$ in the IRAC [3.6] channel, are indicated by open symbols, and those with higher S/N (i.e. more reliable stellar mass) are shown as filled circles. The pink stars indicate the median sizes of galaxies in different stellar mass bins down to our estimated stellar mass limit. The dot-dashed line in each panel represents the mass-size relation for late-type galaxies (i.e., $n<2.5$) in the local universe from \cite{shen2003} and the dashed line represents the best-fit stellar mass-size relation for UV-bright galaxies at $z\sim2$ from \cite{mosleh2011}. According to our simulations, sizes below the dotted lines (corresponding to apparent radii of $0.05''$) have large uncertainties and can be conservatively thought of as upper limits.

The stellar mass-size relation for $z\sim4$ and $z\sim5$ is characterized by $r_{e} \propto M^{\alpha} $ with $\alpha=0.14\pm0.06$ and $0.17\pm0.07$, respectively.  These are significantly different from zero. We note that the correlation coefficients are $0.22$ at $z\sim4$ and $0.37$ at $z\sim5$. At $z\sim6$ there might be a hint of a mass-size relation, but there are too few galaxies in our sample at this redshift to either confirm or rule out a relation. We note that the size and stellar mass of these galaxies need not to be correlated as the sizes are measured in the rest-frame UV and the stellar masses in optical rest-frame. Nonetheless, the results show the persistence of the mass-size relation for star-forming galaxies up to very high redshifts.  We also note that the fraction of contaminants in our dropout samples is relatively small ($<11\%$)(see \cite{bouwens2010, bouwens2007}  for more details) and hence unlikely biased the relations.

The black diamond in the lower left panel of Figure \ref{fig3} shows the position in the mass-size relation of an \textit{i}-dropout galaxy ($SBM03\#1$ in HUDF) from \cite{eyles2005}. This object has a robust spectroscopic redshift ($z=5.83$) and a stellar mass of a few times $10^{10} \msun$. The size estimate for this object is consistent with the median size estimate of i-dropouts in our higher mass bin.

%%%%%%%%%%%%%%%%%%
\subsection{Size Evolution at fixed mass}

\begin{figure*}
\includegraphics[width=\textwidth]{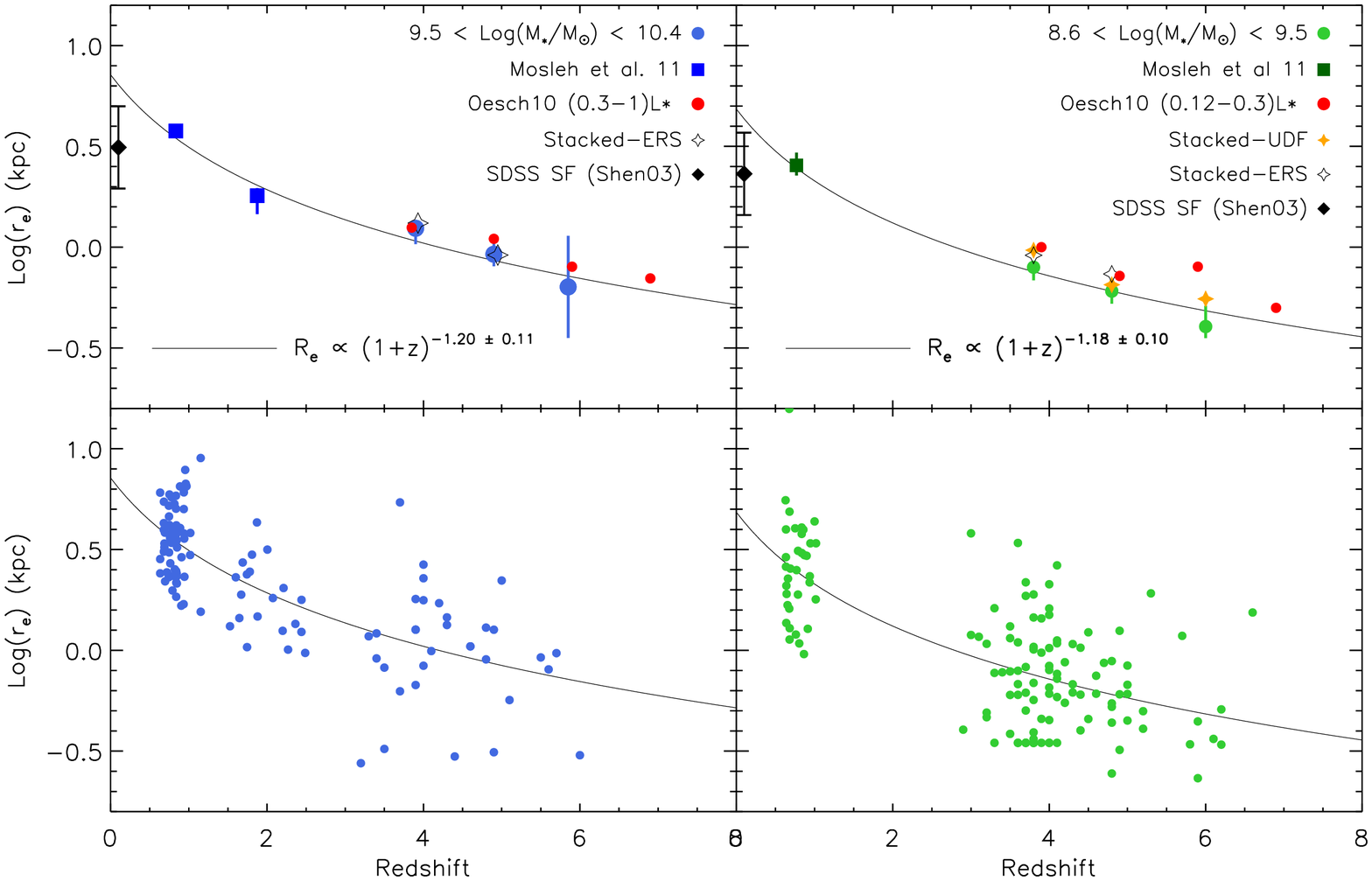}
\caption{The redshift evolution of the sizes of dropout galaxies in two stellar mass ranges: $\log(\mstar/\msun)\sim 9.5-10.4 $ (left panels) and $\log(\mstar/\msun)\sim 8.6-9.5$ (right panels). The lower panels show the distribution of sizes for all sources in our analysis and the sizes of UV-bright galaxies at $z\sim1-3$ within the same mass range from \cite{mosleh2011}. The best power law fits to these data points are shown as solid lines. The red points are from \cite{oesch2010b} for a luminosity-selected sample. The black and yellow stars are the sizes derived by stacking galaxies in the same stellar mass range. The black diamonds in the top panels indicate median size of local late-type galaxies from \cite{shen2003}.}
\label{fig4}
\end{figure*}

Comparing the median sizes of LBGs for a given stellar mass at different epochs, illustrates the evolution of galaxies effective radii at fixed stellar masses with redshift. This size evolution is shown in Figure \ref{fig4}. Galaxies are split into two stellar mass bins: $9.5< \log(\mstar/\msun)<10.4 $ (shown in the left panels) and $8.6 <\log(\mstar/\msun) < 9.5 $ (shown in the right panels). For the objects in the lower redshift range (i.e. $z\sim1-3$), we used sizes measured for UV-bright galaxies in the same stellar mass range from our previous studies by \cite{mosleh2011}. In order to measure the size evolution consistently we use similar technique described in \cite{newman2012} to normalize sizes of galaxies with $r_{e} \propto M^{0.30}$ to stellar mass of $10^{9.7} \msun$ (left panels) and  $10^{9} \msun$ (right panels). In the bottom panels, blue and green points represent normalized half-light radii of dropout objects as a function of redshift. 
 
We fit a simple power law of the form $(1+z)^{-m}$ to the observed points (bottom panels of Figure \ref{fig4}); the best fits are shown as solid lines. In the high stellar mass bin, galaxy sizes are found to evolve as $(1+z)^{-1.20\pm0.11}$, and for galaxies with lower stellar masses the size evolves as $(1+z)^{-1.18\pm0.10}$. 

In the top panels of Figure \ref{fig4}, blue and green solid circles represent the median effective radii of dropouts from this study, and the solid squares are median sizes in lower redshift bins.  The red points are the mean galaxy sizes from \cite{oesch2010b} for two different luminosity ranges:  $(0.3-1)L_{*\ z=3}$ in the left panel and $(0.12-0.3)L_{*\ z=3} $ in the right panel. 

The filled yellow and open black stars in the top panels are sizes based on stacking galaxies from fixed stellar mass ranges for our two fields. For stacking, we used the central positions of objects determined by GALFIT and replaced contaminated pixels from neighbouring sources with sky background values. We then used GALFIT to measure the half-light radii of our final stacked images. As shown by e.g., \cite{hathi2008b} and \cite{oesch2010b}, stacking can reproduce reliable average surface brightness profiles. Thus, the  agreement between the median points and results of stacking suggests that we are not systematically missing light in the extended wings of galaxies and that on average our size estimates are robust.

\begin{figure*}
\includegraphics[width=\textwidth]{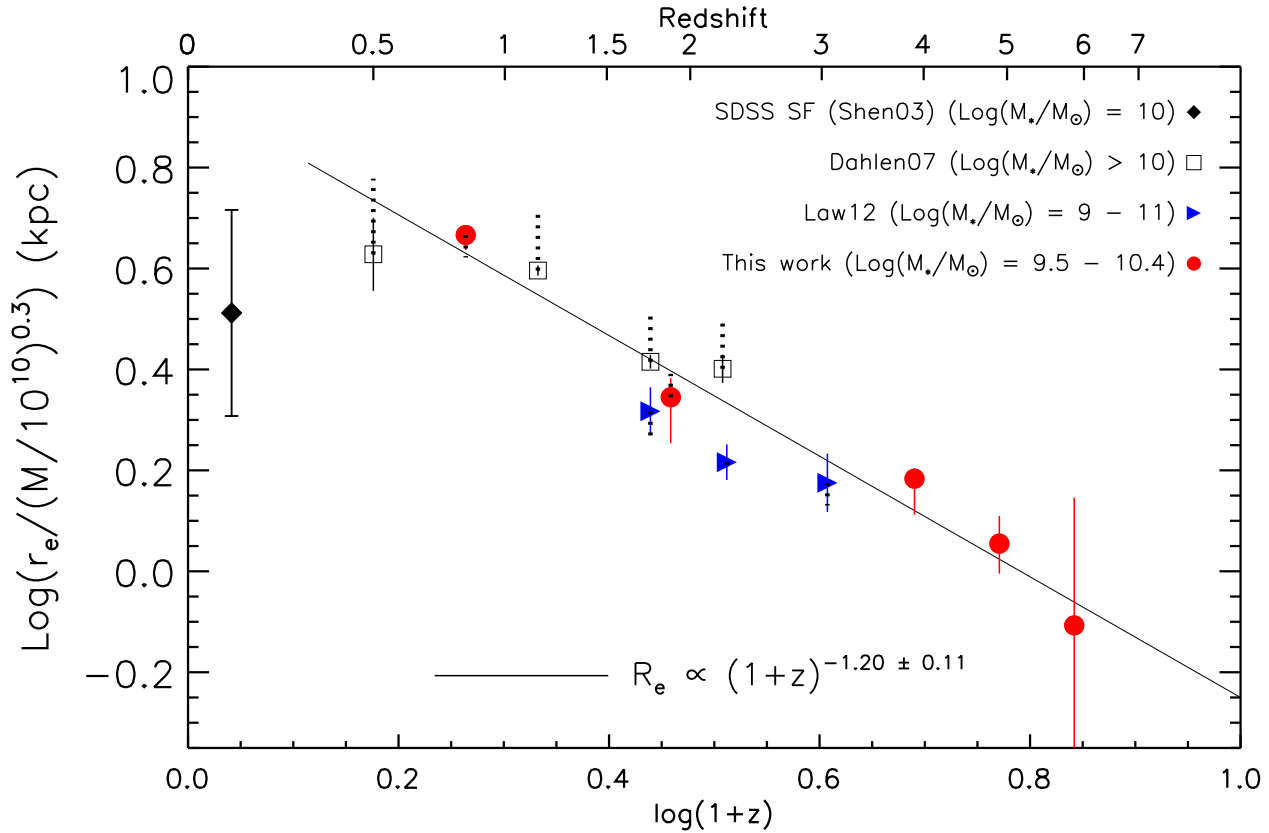}
\caption{Comparison of sizes of UV-bright galaxies measured by different authors. In order to have a consistent comparison, sizes are normalized using $r_{e} \propto M^{0.30}$ for a stellar mass of $10^{10}\msun$. The solid line is the best fit from our analysis ($r_{e} \propto (1+z)^{-1.20\pm0.11}$). The dotted lines represent the original (not normalized) size measurement. The black solid diamond is the median size of late-type galaxies at $z\sim0$ for stellar masses around  $10^{10}\msun$ from \cite{shen2003}, with the measured dispersion shown as error bar. The sizes of UV-bright galaxies at fixed stellar mass increases rapidly towards later cosmic time. However, the sizes measured for local late-type galaxies are smaller than the sizes of UV-bright galaxies at $z\sim 1$. These samples most likely comprise intrinsically different galaxies.} 
\label{fig5}
\end{figure*}

%%%%%%%%%%%%%%%%%%%%%%%%%%%%%%%%%%%%%%%%%%%%%%%%%%%%%%%%%%%%%%%%%%%%%%%%
\section{Summary \& Discussion}

For the first time, we have studied the stellar mass-size relation of LBGs out to $z\sim7$ using ultra-deep WFC3/IR observations in the HUDF and ERS fields. We have shown that the mass-size relation of star-forming galaxies persists to very high redshifts, and that at fixed stellar mass, the sizes of galaxies increase significantly towards later cosmic time. The observed size growth of LBGs studied here - $r_{e}\propto (1+z)^{-1.20\pm0.11}$ for galaxies with stellar mass of $10^{9.5}$-$ 10^{10.4} \msun$ - is in agreement with previous stellar mass-size studies at $z\lesssim 3$ \citep[e.g.,][]{dahlen2007,mosleh2011, nagy2011, law2011}. It is also consistent with the size evolution estimated by other studies based on luminosity-selected samples \citep[e.g.,][]{ferguson2004, bouwens2004, bouwens2006,hathi2008a, oesch2010b}. \\

The redshift dependence of the size evolutions is very similar for both high and low stellar masses. Therefore, the galaxy size evolution might be written as a separable function of mass and redshift and this would be in agreement with  simple models of galaxy formation developed for high redshift systems \citep[see][]{wyithe2011}. However, our sample is not stellar mass complete; hence using deeper samples in future is needed for further investigations.

In Figure \ref{fig5} we compare our estimated size-redshift relation for LBGs with those in \cite{law2011} (blue triangles) and 
\cite{dahlen2007} (open squares). The sizes are normalized to a stellar mass of $10^{10}\msun$. The size estimates from different studies are consistent with the best fit found in this study ($m=1.20\pm0.11$, solid line). This suggests that LBGs may evolve into UV-bright galaxies at $z\sim1$. At $z\sim 6$ these galaxies are extremely compact: $r_{e}\sim0.8$ kpc, at stellar mass of $10^{10}\msun$. However, they grow by a factor of about 6 to $z\sim0.85$. The $z\sim6$ galaxies have a mass size relation close to those of the compact quiescent galaxies at $z\sim2$; For example the normalized size ($r_{e}/(M/10^{10})^{0.3}$ ) of a sample of quiescent galaxies at $z\sim2$ studied by \cite{szomoru2012} is $\sim0.6$ kpc.

The scarcity of observed LBGs between $z\sim0$ and $z\lesssim1$,  complicates the interpretation of the evolution of their stellar mass-size relation to the present time. The median size measured for local late-type (i.e., $n < 2.5$) SDSS galaxies by \cite{shen2003} at the same stellar mass (black solid diamond in Figures \ref{fig4} and \ref{fig5}, corrected to the rest-frame UV  using the analysis by \cite{Gildepaz2007} and \cite{Azzollini2009} and D. Szomoru (private communication)) is smaller than the size of $z\sim 1$ UV-bright galaxies. We note that the SDSS late-type sample is most likely a different galaxy population than the UV-bright sources we study at $z\gtrsim1$, and is therefore not relevant for direct comparison. \cite{overzier2010} studied 30 local Lyman Break Analogs ($z<0.3$), however this sample was selected to have a surface brightness limit and is therefore not characteristic of star-forming galaxies in the nearby universe. In addition, there is some evidence that the evolution of the stellar mass-size relation for star-forming galaxies is slower between $z=1$ and $z=0$ \citep[e.g.,][]{barden2005}. This needs further investigation using homogeneously-selected sample in future.\\ 
  
This work was funded in part by the Marie Curie Initial Training Network ELIXIR of the European Commission under contract PITN-GA-2008-214227.

%%%%%%%%%%%%%%%%%%%%%%%%%%%%%%%%%%%%%%%%%%%%%%%%%%%%%%%%%%%%%%%%%%%%%%%%

\bibliography{apj-jour,Refs}
%%%%%%%%%%%%%%%%%

\end{document}